# Revisit the phase diagram and piezoelectricity of lead zirconate titanate from first principles

Yubai Shi,[1, 2] Ri He,[1, †] Bingwen Zhang,[3] and Zhicheng Zhong[1, 4, ‡]

[1]*CAS Key Laboratory of Magnetic Materials, Devices and Zhejiang Province Key Laboratory of Magnetic Materials and Application Technology, Ningbo Institute of Materials Technology and Engineering, Chinese Academy of Sciences, Ningbo 315201, China*

[2]*College of Materials Science and Opto-Electronic Technology, University of Chinese Academy of Sciences, Beijing 100049, China*

[3]*Fujian Key Laboratory of Functional Marine Sensing Materials, Center for Advanced Marine Materials and Smart Sensors, College of Material and Chemical Engineering, Minjiang University, Fuzhou 350108, China*

[4]*China Center of Materials Science and Optoelectronics Engineering, University of Chinese Academy of Sciences, Beijing 100049, China*

Lead zirconate titanate (PbZr$_{1-x}$Ti$_x$O$_3$, PZT) exhibits excellent piezoelectric properties in the morphotropic phase boundary (MPB) region of its temperature-composition phase diagram. However, the microscopic origin of its high piezoelectric response remains controversial. Here, we develop a machine-learning-based deep potential (DP) model of PZT using the training dataset from first principles density functional theory calculations. Based on DP-assisted large-scale atomic simulations, we reproduce the temperature-composition phase diagram of PZT, in good agreement with the experiment except the absence of structural transition from *R3c* to *R3m*. We find that the rhombohedral phase maintains *R3c* symmetry with slight oxygen octahedral tilting as increase of temperature, instead of appearing *R3m* symmetry. This discrepancy can trace back to the lack of experimental measurements to identify such slight octahedral tilting. More importantly, we clarify the atomic-level feature of PZT at the MPB, exhibiting the competing coupling of ferroelectric nanodomains with various polarization orientations. The high piezoelectric response is driven by polarization rotation of nanodomains induced by an external electric field.

---

[†]heri@nimte.ac.cn

[‡]zhong@nimte.ac.cn

## I. INTRODUCTION

Lead zirconate titanate (PbZr$_{1-x}$Ti$_x$O$_3$, PZT) is of great interest due to its excellent piezoelectric response. PZT possesses the typical ABO$_3$ perovskite structure, where oxygen atoms form octahedra, Pb atoms occupy the interstices of the octahedra, and Zr/Ti atoms reside at the center of the octahedra with disordered distribution. Thus, the solid solution structure leads to a complex temperature-composition phase diagram[1,2]. Especially, the nearly vertical phase boundary between the tetragonal and rhombohedral regions of the phase diagram close to $x = 0.5$, called morphotropic phase boundary (MPB), exhibits fascinating physical properties [3,4].

Although it is well known that the high piezoelectricity of PZT is related to MPB, the exact microscopic origins is still unclear. The MPB solid solution presents complex multi-domain structures, featuring the coexistence and coupling competition of multiple phases, which contains the rhombohedral and tetragonal phase with polarization along the <111> and <001> orientations, respectively. Therefore, the study of its properties is rather complicated. Various theories have been proposed to explain the origins of high piezoelectricity origins at the MPB. For example, Isupov et al. believed the mechanism resulting in enhanced properties was due to the coexistence of tetragonal and rhombohedral phases in the vicinity of the MPB and Kakegawa et al. confirmed the coexistence of these phases by experiments [5-10]. Furthermore, Fu et al. suggested that a large piezoelectric response was driven by polarization rotation induced by an external field [11-14], whereas Frantti et al. proposed that the phase instability, in contrast to the polarization rotation, was responsible for the large piezoelectric properties observed in PZT near the MPB [15-18]. Besides, Li et al. found that local fluctuations of a polarization vector might result in a unique domain structure, such as nanometer ferroelectric domains, which could be the crucial factor responsible for the excellent piezoelectric response in PZT [19-23]. More recently, Yan et al. believed that an oxygen octahedral tilting/untilting transition at *R3c/P4mm* MPB leads to a high free energy barrier and results in low piezoelectric activity, whereas *R3m/P4mm* MPB with octahedral untilting produces large piezoelectricity due to low free energy barrier at such MPB [24]. Due to the experimental challenges in characterizing the composition distribution and complex microscopic structures at the MPB, resolving these controversies is rather difficult.

Atomic simulation from first principles is an effective method for studying such complex solid solutions and its thermodynamics properties. Regrettably, structural thermodynamics properties of these complex systems prove to be beyond the capacity of DFT due to the large system sizes and simulation times involved. To solve this issue, we utilize machine-learning-based deep potential (DP) model to perform large-scale atomic simulation. Machine-learning potentials use a deep neural network to capture the potential energy surface of material system in DFT accuracy. Recent research utilizing DP method has witnessed numerous breakthroughs in the field of ferroelectric perovskite materials [25-28], providing compelling evidence for the reliability of this technique.

Herein, we develop a machine-learning-based potential for PZT using dataset from DFT calculations. The DP model can effectively reproduce the temperature-composition phase diagram and properties of PZT. Importantly, DP predicted results reveal the absence of the *R3m* structure, which was claimed to be observed in experiment. We attribute this discrepancy to limitations in experimental measurements. We calculated the piezoelectric constant ($d_{33}$) over a whole range of compositions, demonstrating highest piezoelectric response of the MPB region compare to other components. Furthermore, we reveal the mechanism underlying the ultrahigh piezoelectric response of MPB is driven by polarization rotation. Our work clarifies the origin of high piezoelectricity at the MPB in PZT from the atomic level, paving the way for the experimental design of novel piezoelectric materials.

## II. COMPUTATIONAL METHODS
### A. Generation of DP model

The selection of the training datasets is crucial for the accuracy of the DP model. This requires the training datasets to comprehensively cover the potential energy surface of system. Generating configurations for the training set is not a trivial task. Here, we utilize DP-DEN to generate training datasets encompassing a sufficiently broad space of relevant configurations [29]. DP-GEN has been widely utilized over the past years, and DP model for variation material systems have been developed by this strategy [30]. Here, we emphasize only a few key points of the training process. DP-GEN is a concurrent learning procedure, and the workflow of each iteration includes three main steps: training, exploration, and labeling (see Fig. 1). Considering that we aim to describe properties of PbZr$_x$Ti$_{1-x}$O$_3$ with continuous variations of $x$, the training datasets include different $x$ values: 0, 0.125, 0.25, 0.5, 0.75, 0.875 and 1. When describing the properties of a system as it varies with temperature, it is necessary to consider different initial structures at different temperatures. For example, when $x = 0$, there are both $Pm\bar{3}m$ cubic phase which stable at high temperatures and $Pbam$ orthorhombic phase which stable at low temperatures. The configurations of the initial datasets are generated by perturbing atomic coordinates of these structures. Starting with the above datasets including configurations and corresponding density DFT calculated energies, four different DP models are trained using the Deep molecule dynamic simulations (DPMD) method [31], based on different values of deep neural network parameters. The exploration step is performed in which one of the DP models is used for molecule dynamic simulations (MD) at given pressure and temperature to explore the configuration space. We established an extensive exploration range spanning from 50 to 1000 K and from -100 to $10^5$ Pa. For all sampled configurations in MD trajectories, the other three DP models will predict the atomic forces of all atoms. The maximum standard deviation of atomic forces $F_i$ is used to construct an error indicator for labeling: $\sigma_f^{max} = max \sqrt{\langle |F_i - \langle F_i \rangle|^2 \rangle}$, where $\langle ... \rangle$ indicates the average of DP-predicted force, and $F_i$ denotes the predicted force on the atom $i$. When $\sigma_f^{max} > \sigma_{high}$, the corresponding configuration is labeled as a failure. Only configurations with $\sigma_{low} < \sigma_f^{max} < \sigma_{high}$ undergo DFT calculations, then get added to the next iteration's training dataset. Here, the exploration of each system is considered converged when the percentage of accurate configurations ($\sigma_{low} < \sigma_f^{max}$) is > 99%. It's worth noting that all systems are not iterated together. Instead, systems with different $x$ values are individually iterated to ensure each reaches the convergence accuracy. The DP-GEN process stops when all systems satisfy converged. Finally, all datasets are collected and combined into the final training dataset for a long-step training. This approach offers the advantage of improving efficiency. What's more, the DP model trained this way can describe the properties of PZT in continuous variations of $x$ values, and it possesses the same level of accuracy as individually training for a particular $x$ value. We also utilize the compressed model in this work. This method has been proven to significantly enhance computational efficiency of DPMD, with negligible loss in accuracy [32]. Long-range electrostatic interactions play an important role in dielectric materials simulation. Introducing a cutoff radius limits the interaction range and may potentially overlook certain long-range effects. Nevertheless, in many cases, the finite-range model indeed gives an accuracy of ~1 meV/atom in energy, which is comparable with the intrinsic error of the functional approximation adopted in DFT [26,33-35]. This level of accuracy is sufficient for most physical properties of practical interest. Furthermore, a recent method proves that DP model is capable of accurately describing long-range electrostatic interactions [36]. Thus, it is possible to simulate piezoelectric materials with long-range electrostatic interactions using the DP model in the future.

### B. DFT calculations

The quality of the initial datasets determines the accuracy of the ultimately trained DP model. The initial datasets required for training a deep neural network are entirely generated by the Vienna *Ab initio* Simulation Package (VASP) [37,38], which maintains the precision of first principles. All DFT calculations are performed using a plane-wave basis set with a cutoff energy of 600 eV, and the exchange-correlation energy of electrons is described in the generalized gradient approximation using the PBEsol functional [39]. The Brillouin zone is sampled using K-points with a minimum spacing of 0.2 Å$^{-1}$, including the gamma point. The calculation of the phonon spectrum was carried out using the Phonopy package [40,41], and the force constant matrix was generated using the finite displacement method, with a $2 \times 2 \times 2$ supercell. Additionally, we introduce the non-analytic correction (NAC) when performing calculations for the cubic structure to remove the degeneracy of longitudinal optical and transverse optical branches.

## C. MD simulations

MD simulations starting with a $10 \times 10 \times 10$ supercell (5000 atoms) are performed by LAMMPS code with periodic boundary conditions and at standard pressure to model the temperature and composition-driven structural transition of PZT [42,43]. Simulations using a $20 \times 20 \times 20$ supercell can give similar results. MD simulations adopt the isobaric-isothermal (NPT) ensemble with temperature set from 100 to 700 K. The time step is set to 0.001 ps. At a specified temperature, the equilibrium run is 40 ps. Running for 100 ps near the phase transition point to ensure an adequate amount of statistical data. The simulation environment is set as a triclinic box, allowing for full relaxation of lattice constants and lattice angles. In PZT solid solution, Zr and Ti exhibit a disordered arrangement. Our tests indicate that different disordered arrangements have negligible impact on energy with 0.03 meV/f.u.. Applying an electric field to the system is achieved by applying an external force to the atoms as an equivalent method. The forces acting on different atoms are determined by multiplying the electric field by the reference Born effective charge [44]: $Z^*_{Pb} = 3.90$, $Z^*_{Zr} = 5.85$, $Z^*_{Ti} = 7.06$. Taking into account the solid solution characteristics of PZT, a dynamic average adjustment scheme is applied to the Born effective charge of oxygen atoms: $Z^*_O = -[Z^*_{Pb} + (1-x)Z^*_{Ti} + xZ^*_{Zr}]/3$, where $x$ represents the concentration of Zr. The piezoelectric coefficient $d_{33}$ is obtained based on: $d_{33} = (\partial x_3/\partial E_3)_X$, where $E_3$ is external electric field, $x_3$ the strain induced, 3 the along the $z$ axis, and $X$ represents the external stress. The polarization orientation diagram is based on the quasi-two-dimensional structure of $100 \times 100 \times 2$ supercell obtained after high-temperature annealing at 700 K.

## III. RESULTS AND DISCUSSIONS

### A. Accuracy of DP model

Evaluating the accuracy of the DP model is essential to ensure the reliability of the DPMD simulation results. We compare the energies and atomic forces calculated using the DP model and DFT calculations for the all configurations in the final training dataset. As shown in Fig. 2(a) and (b), the mean absolute errors of energy ($\Delta E =$

$|E_{DFT} - E_{DP}|$) and atomic force ($\Delta F = |F_{DFT} - F_{DP}|$) between DP and DFT are 0.86 meV/atom and 0.04 eV/Å, respectively. It is evident that the DP model well reproduces DFT results of PZT in different components. These demonstrate the superior representability and flexibility of DP that a single DP model can describe the high-dimensional energy function of complex solid solutions composed of different components of two materials. Due to force constant represents the second derivative of energy with respect to atomic displacement, it serves as a stricter criterion for testing the accuracy of DP model. Therefore, we employ the DP model and DFT calculations to determine the phonon dispersion relations of PTO. As shown in Fig. 2(c), the DP model calculated phonon dispersion of cubic PTO agrees well with the DFT result. It is clear that imaginary frequencies appear at the R and Γ points，indicating cubic phase has multiple lattice instability at 0 K. For the tetragonal phase, DP model correctly predicted its dynamically stable vibrational modes over the whole Brillouin zone [see Fig. 2(d)]. These results demonstrate that the DP model can effectively describe the characteristics of the interaction between atoms.

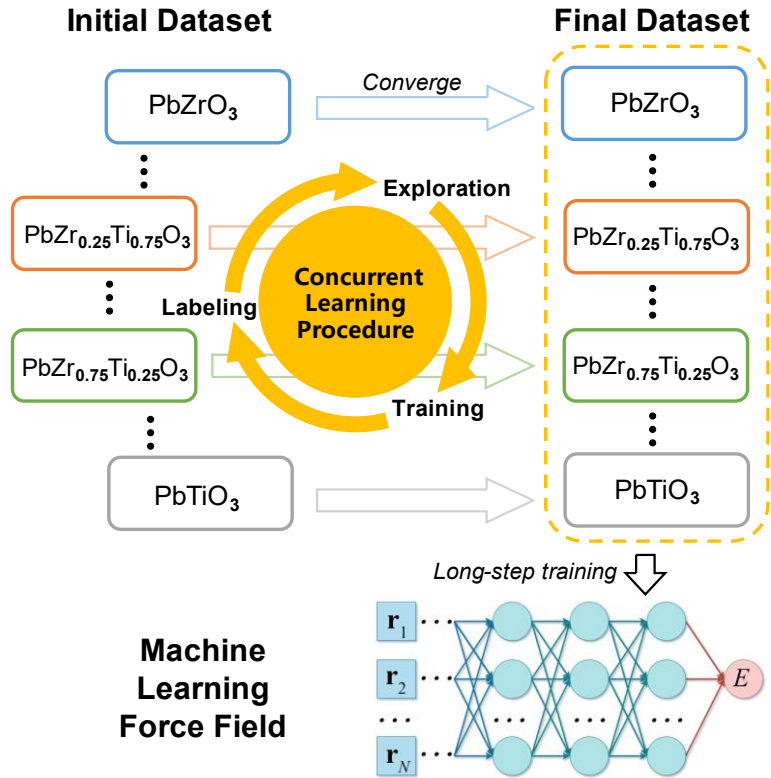

FIG. 1. The workflow of the DP model generation for PbZr$_x$Ti$_{1-x}$O$_3$. Iterative training with different components individually, including $x = 0, 0.125, 0.25, 0.5, 0.75, 0.875$, and $1$. After convergence of each component, the datasets with different $x$ are collected into the final dataset. A DP model was generated through a long-step training based on this final training dataset.

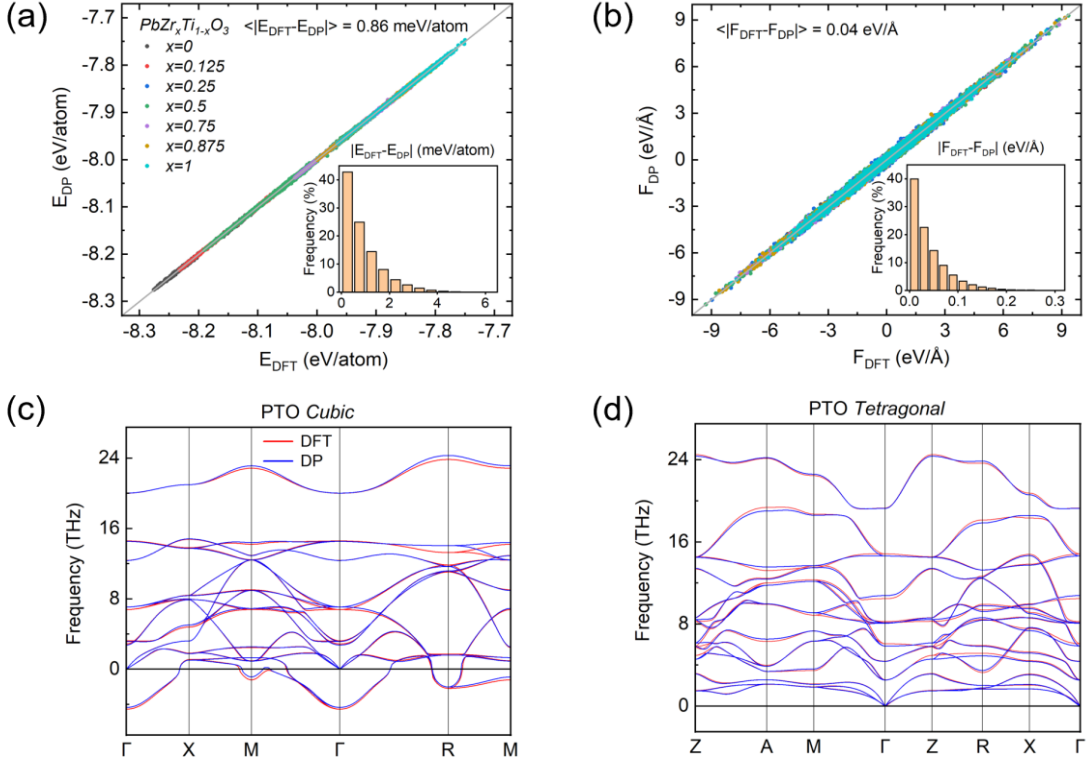

FIG. 2. The comparative analysis involves assessing (a) total energies and (b) atomic forces of PZT with different $x$ values, employing both the DP predictions and DFT calculations. Phonon spectra of (c) tetragonal PTO, and (d) cubic PTO.

## B. Phase diagram of PZT solid solutions

In the temperature-composition phase diagram of PZT, there exists both compositional disorder and coupling competition between multiple structural phases. The phase diagram is readily reproducible in experiments and has been extensively studied since its initial depiction in 1971 by Jaffe et al. [4]. But relevant theoretical researches are still lacking due to its higher computational precision requirements and structural complexity [45]. Here, the DP model allow us to perform DPMD simulations for different compositions at finite temperatures. We present the temperature phase diagrams for pure PZO and PTO in Fig. 3(a) and (b). It show that PZO transitions from the antiferroelectric orthorhombic (O) phase to the cubic (C) phase at 460 K, while PTO transitions from the ferroelectric tetragonal (T) phase to the cubic (C) phase at 530 K, in consistent with the experiment works [46]. In Fig. 3(c) and (d), we plot the temperature phase diagram at $x = 0.5$. It is found that PZT successively exhibits four phases: rhombohedral (R), monoclinic (M), tetragonal (T), cubic (C), as the temperature rises. The presence of the monoclinic phase between the rhombohedral phase and tetragonal phase at 370 K~470 K, which differs from the one reported in a recent work [35]. Their calculation results indicate a direct transition from the rhombohedral phase to the tetragonal phase without bridged monoclinic phase. We assume the reason for the difference is that the universal model for perovskite oxides might lose the accuracy to describe minor details within certain systems. We also obtain the temperature-composition phase diagram over a whole range of compositions shown in Fig. 4. It is found five phases in the phase diagram, consistent with experiment. The antiferroelectric orthorhombic phase is observed in the range of $x = 0$~$0.08$, on the left of blue line. As the $x$ increasing, the ferroelectric rhombohedral phase with polarization along the [111] direction appears. Furthermore, when $x$ closes to ~0.5, there comes the most important region in the phase diagram called MPB, marked by red line. On the right side of MPB, the most stable structure is tetragonal phase with polarization along the [001] direction. Additionally, we find the monoclinic phase at the MPB, indicating the monoclinic phase is a bridge between the rhombohedral phase and the tetragonal phase, in agreement with previous experiment [47,48]. At high temperatures, the ferroelectric phases transform to paraelectric cubic phase in a whole range of compositions as show in Fig. 4. A detail of *R3m* and *R3c* structures in phase diagram will be discussed below. The results of the phase diagram agree with the previous experiment results [4]. However, The DP-predicted phase diagram suggests a slight decrease in ferroelectric-paraelectric transition temperature on the side close to PTO. Experimentally, the temperature monotonously increases with $x$. This issue also appeared in other DP research [35], and it does not affect the investigation of the MPB region.

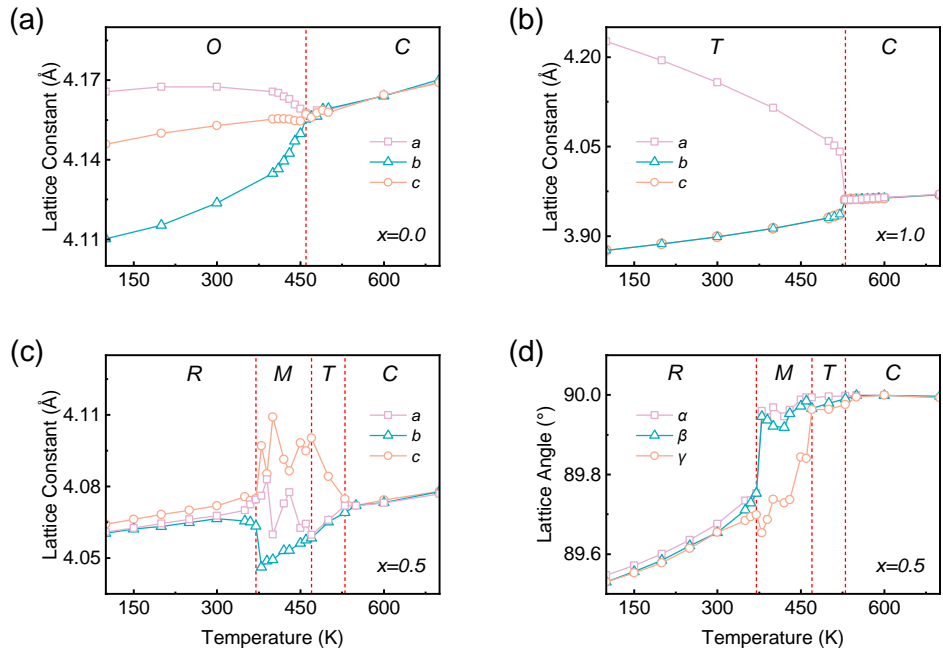

FIG. 3. Schematic representation of the temperature-dependent lattice constants for (a) $PbZrO_3$, (b) $PbTiO_3$, (c) $PbZr_{0.5}Ti_{0.5}O_3$ and (d) lattice angles for $PbZr_{0.5}Ti_{0.5}O_3$. O, R, T, M, and C represent the orthorhombic, rhombohedral, tetragonal, monoclinic, and cubic phases, respectively.

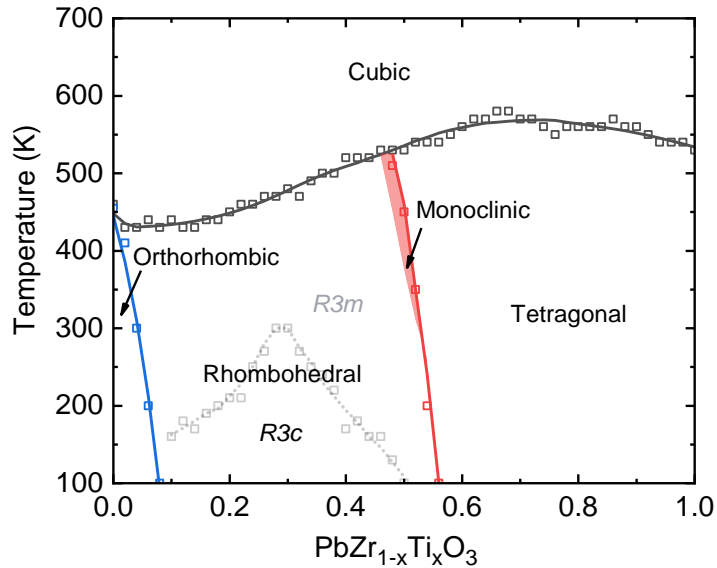

FIG. 4. DP predicted temperature-composition phase diagram of PZT. Red line represents the MPB and the red shaded area represents the monoclinic phase near the MPB. The gray dashed line represents the artificially defined hypothetical boundary between the *R3m* and *R3c* structures.

According to the previous experiments [49,50], both the *R3c* and *R3m* symmetries exist in the rhombohedral phase region. researchers generally believe that there exists a structure transition from *R3c* to *R3m* with increasing temperature. However, our calculated results indicate the absence of *R3m* symmetry. The distinction between these two symmetries is depicted in Fig. 5(a): the *R3c* symmetry can be characterized by the neighboring $TiO_6$ octahedra tilting with a small angle in the opposite directions along the [111] direction, whereas the $TiO_6$ octahedra do not tilt in *R3m* symmetry. The periodically arranged oxygen atoms in the first and second octahedra are labeled as O1 (colored blue) and O2 (colored orange) along the [111] direction. We perform statistical analysis of the rotation angles $\theta$ between O1 and O2 versus temperature for different values of *x*. The result in Fig. 5(b) shows that $\theta$ decreases to zero approximately only because it turns to cubic phase at high temperature directly, around 500 K, instead of transitioning to *R3m*.

We assume the structure transition which was observed in rhombohedral phase in experimental data is a result of the limitation of experimental measurements. The real-space distribution schematic diagram of O1 and O2 is provided in Fig. 5(c)-(e). We performed a statistical analysis to determine the frequency of atom occurrence at various spatial positions. When O1 and O2 are counted separately, it is clear that O1 and O2 exhibit a Gaussian distribution at their respective positions. Considering that experimentally distinguishing between oxygen atoms in different octahedra is not feasible, we have combined O1 and O2 for statistical purposes. The results indicate that with the increase of temperature, the two distribution peaks gradually merge into one single peak, corresponding to the inability to distinguish between O1 and O2 in real-space. Therefore, some previous studies have suggested the presence of a structural transition from *R3c* to *R3m*, which actually does not occur. We artificially define that if the height of saddle point exceeds the half height of the two sides peaks, it is "considered" to transfer to the structure in *R3m* symmetry. For example, as shown in Fig. 5(d), at *x* = 0.4 and T = 180 K, the merged peak of O1 and O2 reach the "distinguishing limit". According to this criterion, we mark the gray dash line in Fig. 3, fitting well with the previous studies.

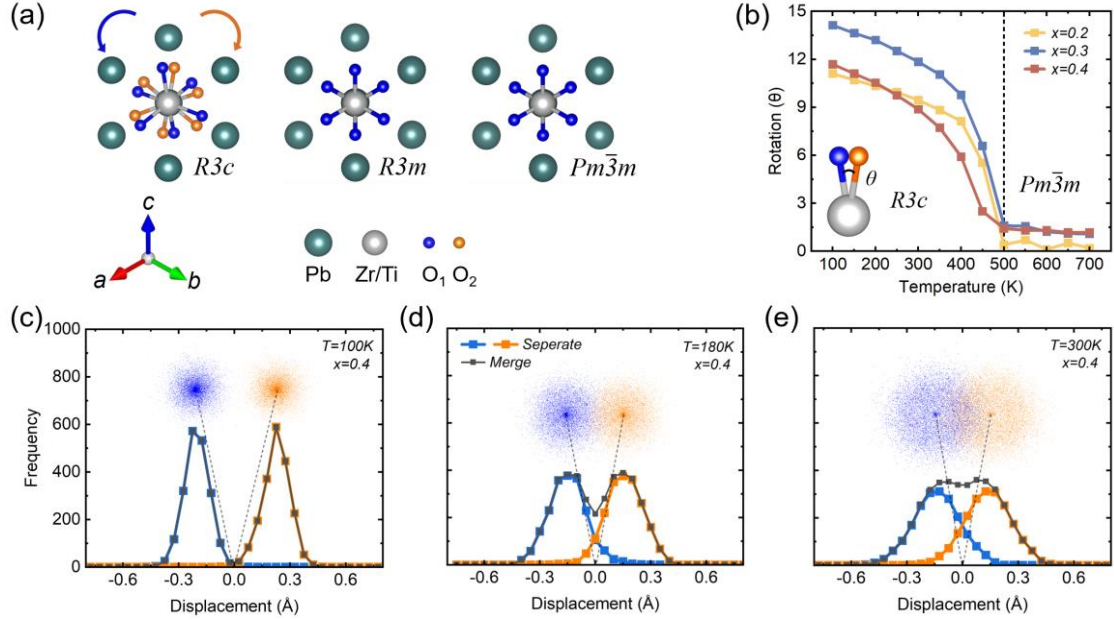

FIG. 5. (a) Crystal structures along the [111] direction in the *R3c*, *R3m*, and *Pm$\bar{3}$m* symmetries. O1 and O2 represent the oxygen atoms in the neighbor octahedra with reverse-tilting. (b) The angle variation between O1 and O2 as a function of temperature. When $x = 0.4$, the real-space distribution of O1 and O2 at T = (c) 100 K, (d) 180 K, (e) 300 K.

## C. High piezoelectric performance in MPB

The next issue is to determine exact origin of high piezoelectricity at the MPB region. Piezoelectric materials exhibit a strain response under an electric field. The piezoelectric coefficient ($d_{33}$) can be characterized by the differential of strain over electric field. In Fig. 6(a) and (b), we plot the strain-electric field curves near the MPB region at 100 K and 300 K. It clearly shows that the curve exhibits different behaviors with different *x*. Take the 100 K as an example, when $x \lesssim 0.50$, strain increases linearly with the electric field. When $0.5 < x \lesssim 0.56$, the strain exhibits three stages: linear, abrupt change, and linear. The abrupt change is due to the polarization rotation, and details will be discussed below. With increasing *x*, the electric field strength required for the abrupt change decreases. When $x > 0.56$, the relationship between strain and electric field reverts back to being linear. We define $x = 0.56$ as the critical point at 100 K. As *x* exceeds this value, no sudden change occurs in strain. Similar behavior is observed at 300 K as well, with a different critical point occurring at $x = 0.52$. Based the derivative of strain with respect to electric field with range of 0~1($\times 10^2$ kV/cm), we extract the piezoelectric coefficient $d_{33}$ as shown in Fig. 6(c) (The detailed calculation content of $d_{33}$ can be seen in the method). Evidently, $d_{33}$ experiences a significant increase to its maximum at the critical points, where $x = 0.56$, T = 100 K, and $x = 0.52$, T = 300 K. The trend of $d_{33}$ changing with $x$ aligns well with previous research [3]. The conclusion drawn is that critical points are highly temperature-sensitive with $x$ decreasing as temperature increases. Its trajectory aligns closely with the red line in Fig. 4, corresponding to the MPB in the phase diagram. Thus, MPB is indeed the region with the best piezoelectric performance. The DP-predicted $d_{33}$ are larger than the experiment results [51]. We attribute this phenomenon to the order degree of Zr/Ti atomic arrangement. Extensive details will be elaborated in next publication.

To unveil the microscale mechanisms behind the high piezoelectric properties in the MPB region, we plot the snapshots of electric dipole configurations with different electric field at $x = 0.5$, T = 300 K in Fig. 7. When the electric field is zero, the dipoles exhibit a complex nanodomain structure with polarization orientations roughly along the eight directions, including [111] and its equivalent directions. With increasing electric field strength to $1\times 10^2$ kV/cm, an increasing number of dipoles start to reorient towards the [001] direction, and nanodomains with polarization along [001] gradually grow larger. As increasing to $2\times 10^2$ kV/cm, almost all dipoles align exclusively in the [001] direction, representing the polarization rotation tends towards saturation. The polarization rotation process corresponds to the abrupt change stage in the strain-electric field curve of Fig. 6(a), also corresponding to the phase transition from rhombohedral phase to tetragonal phase of PZT. Moreover, The intermediate transitional monoclinic phase observed in phase diagram suggests that there is metastable [011] orientation between the polarizations in the [111] and [001] directions, Serving as a bridge for the phase transition [11]. As the electric field further increase to $3\times 10^2$ kV/cm, dipoles aligning towards the [001] direction only increase in magnitude, corresponding to the linear stage on the curve. The process of polarization rotation leads to the high $d_{33}$ value of 2263 pC/N at the MPB. We also plotted snapshots of pure PTO for comparison of piezoelectricity. As shown in Fig. 8, PTO exhibits a tetragonal phase with different polarization orientations, resulting in a simpler domain structure. Unlike at the MPB, the domain structure evolution shows minimal changes with the increasing electric field instead of polarization rotation. This corresponds to the strain-electric field curve that linearly ascends with a small slope. The calculated $d_{33}$ is 471 pC/N, significantly lower than the result at the MPB. Therefore, a small electric field can induce the polarization rotation as $x$ is close to the MPB, leading to a significant change in lattice constants, which explains the outstanding piezoelectricity of PZT at the MPB.

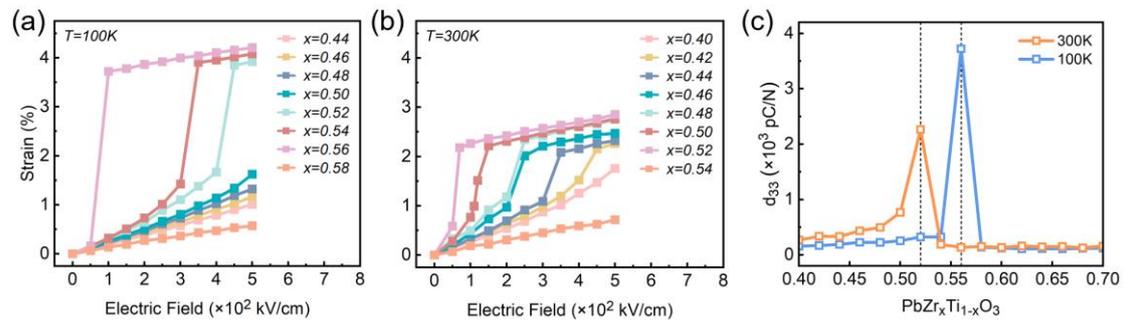

FIG. 6. Strain variation with electric field at (a) 100 K and (b) 300 K. Different colors represent distinct values of $x$. A higher slope indicates better piezoelectric performance. (c) The variation of $d_{33}$ with respect to $x$ at different temperatures. $d_{33}$ reach to its maximum at $x = 0.56$, T = 100 K, and $x = 0.52$, T = 300 K.

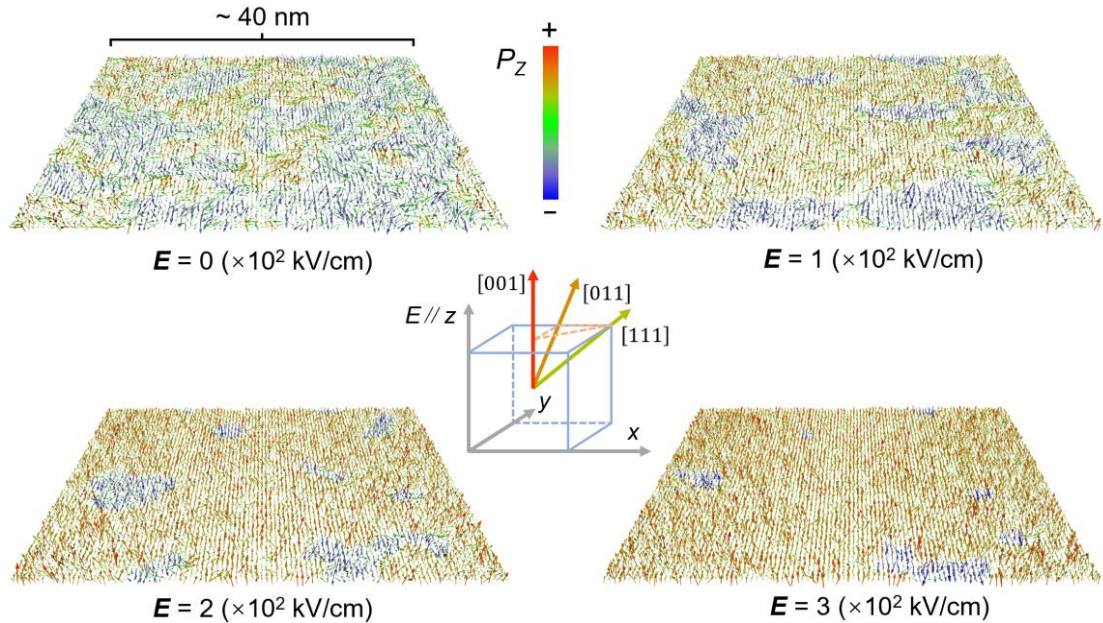

FIG. 7. Typical snapshots of dipole configurations of PbZr$_{0.5}$Ti$_{0.5}$O$_3$ with different electric field at 300 K. The color bar represents the polarization component along the z-direction. Each arrow

represents the local electric dipole vector within a pseudocubic unit cell. Schematic diagram of polarization orientations for the rhombohedral phase along [111], the monoclinic phase along [011], and the tetragonal phase along [001], in the center of the Figure.

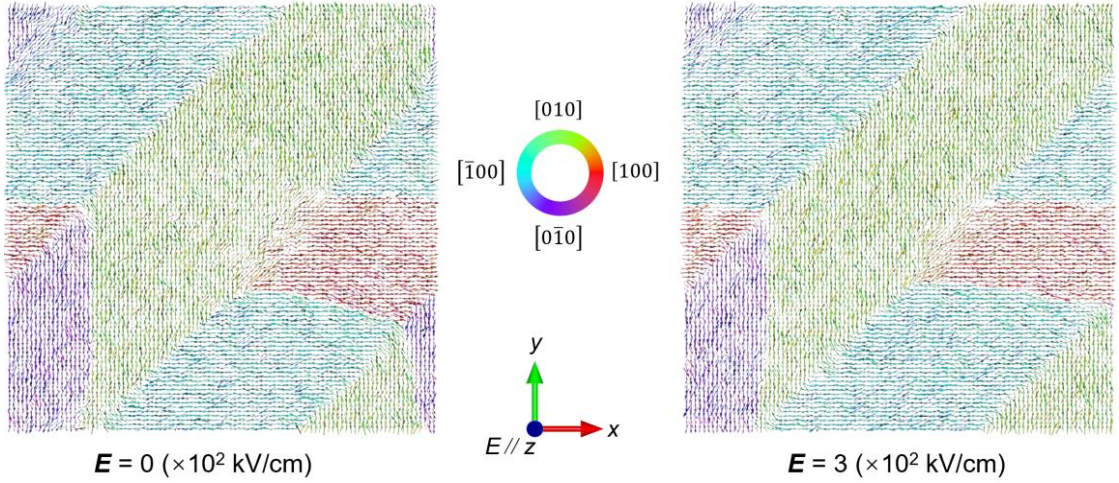

FIG. 8. Typical snapshots of dipole configurations of PTO with different electric field at 300 K. The color circle represents the polarization orientation. Each arrow represents the local electric dipole vector within a pseudocubic unit cell. The domain structure exhibits a tiny response with the electric field.

## IV. CONCLUSION

In summary, we develop the DP model based on the machine learning method to simulate the piezoelectric property and phase transition of PZT, particularly near the MPB. Initially, we reproduce the temperature-composition phase diagram of PZT by computational methods. More importantly, we highlight the absence of the *R3m* symmetry in rhombohedral phase, contrary to experimental observation. *R3m* and *R3c* symmetries are very similar, except for oxygen octahedral tilting along [111] direction in *R3c*. We find that the rhombohedral phase maintains *R3c* symmetry with slight oxygen octahedral tilting as increase of temperature, instead of appearing *R3m* symmetry. The oxygen octahedral tilting becomes so small that experimental measurements are hard to distinguish. We extensively discuss the piezoelectric properties near the MPB. In contrast to the simple domain structure of pure PTO, PZT at the MPB exhibits competition of complex ferroelectric nanodomain structures with different polarization orientations. This clarifies the critical role of electric field-driven polarization rotation in the high piezoelectricity phenomenon. Our work demonstrate that the DP model can accurately captures the kinetic and thermodynamic properties of complex solid solution domain structures at the atomic-level, crucial for understanding the microscopic origins of piezoelectricity and exploring the realm of dielectric energy storage.

## ACKNOWLEDGMENTS

This work was supported by the National Key R&D Program of China (Grants No. 2021YFA0718900, No. 2022YFA1403000, No. 2021YFE0194200), the Key Research Program of Frontier Sciences of CAS (Grant No. ZDBS-LY-SLH008), the National Nature Science Foundation of China (Grants No. 11974365, No. 12204496, No 12161141015), the K.C. Wong Education Foundation (Grant No. GJTD-2020-11), and the Science Center of the National Science Foundation of China (Grant No. 52088101). The authors thank Jiyuan Yang from Westlake University for sharing the phonon spectrums of PTO.

**Reference**

[1] K. Uchino, *Ferroelectric devices* (CRC press, 2018).
[2] N. Izyumskaya, Y.-I. Alivov, S.-J. Cho, H. Morkoç, H. Lee, and Y.-S. Kang, Processing, structure, properties, and applications of PZT thin films, Critical reviews in solid state and materials sciences **32**, 111 (2007).
[3] B. Jaffe, R. Roth, and S. Marzullo, Piezoelectric properties of lead zirconate‐lead titanate solid‐solution ceramics, Journal of Applied Physics **25**, 809 (1954).
[4] B. Jaffe, W. R. Cook, and H. L. Jaffe, *Piezoelectric Ceramics* (Academic Press, 1971).
[5] V. Isupov, Dielectric polarization of PbTiO3-PbZrO3 solid solutions, Soviet Physics Solid State, USSR **12**, 1084 (1970).
[6] P. Ari-Gur and L. Benguigui, X-ray study of the PZT solid solutions near the morphotropic phase transition, Solid State Communications **15**, 1077 (1974).
[7] P. Ari-Gur and L. Benguigui, Direct determination of the coexistence region in the solid solutions Pb (ZrxTi1-x) O3, Journal of Physics D: Applied Physics **8**, 1856 (1975).
[8] V. Isupov, Comments on the paper "X-ray study of the PZT solid solutions near the morphotropic phase transition", Solid State Communications **17**, 1331 (1975).
[9] K. Kakegawa, J. Mohri, T. Takahashi, H. Yamamura, and S. Shirasaki, A compositional fluctuation and properties of Pb (Zr, Ti) O3, Solid State Communications **24**, 769 (1977).
[10] S. M. Gupta and D. Viehland, Tetragonal to rhombohedral transformation in the

lead zirconium titanate lead magnesium niobate-lead titanate crystalline solution, Journal of applied physics **83**, 407 (1998).

[11] H. Fu and R. E. Cohen, Polarization rotation mechanism for ultrahigh electromechanical response in single-crystal piezoelectrics, Nature **403**, 281 (2000).

[12] R. Guo, L. Cross, S. Park, B. Noheda, D. Cox, and G. Shirane, Origin of the high piezoelectric response in PbZr 1− x Ti x O 3, Physical Review Letters **84**, 5423 (2000).

[13] L. Bellaiche and D. Vanderbilt, Intrinsic piezoelectric response in perovskite alloys: PMN-PT versus PZT, Physical review letters **83**, 1347 (1999).

[14] M. Ahart, M. Somayazulu, R. Cohen, P. Ganesh, P. Dera, H.-k. Mao, R. J. Hemley, Y. Ren, P. Liermann, and Z. Wu, Origin of morphotropic phase boundaries in ferroelectrics, Nature **451**, 545 (2008).

[15] J. Frantti, J. Lappalainen, S. Eriksson, S. Ivanov, V. Lantto, S. Nishio, M. Kakihana, and H. Rundlöf, Neutron diffraction studies of Pb (ZrxTi1− x) O3 ceramics, Ferroelectrics **261**, 193 (2001).

[16] J. Frantti, S. Ivanov, J. Lappalainen, S. Eriksson, V. Lantto, S. Nishio, M. Kakihana, and H. Rundlöf, Local and average structure of lead titanate based ceramics, Ferroelectrics **266**, 409 (2002).

[17] J. Frantti, Y. Fujioka, and R. Nieminen, Pressure-induced phase transitions in PbTiO3: a query for the polarization rotation theory, The Journal of Physical Chemistry B **111**, 4287 (2007).

[18] J. Frantti, Notes of the recent structural studies on lead zirconate titanate, The Journal of Physical Chemistry B **112**, 6521 (2008).

[19] J. Li, R. Rogan, E. Üstündag, and K. Bhattacharya, Domain switching in polycrystalline ferroelectric ceramics, Nature materials **4**, 776 (2005).

[20] T. Asada and Y. Koyama, Ferroelectric domain structures around the morphotropic phase boundary of the piezoelectric material Pb Zr 1− x Ti x O 3, Physical Review B **75**, 214111 (2007).

[21] Y. Jin, Y. U. Wang, A. G. Khachaturyan, J. Li, and D. Viehland, Conformal miniaturization of domains with low domain-wall energy: Monoclinic ferroelectric states near the morphotropic phase boundaries, Physical Review Letters **91**, 197601 (2003).

[22] K. A. Schönau, L. A. Schmitt, M. Knapp, H. Fuess, R.-A. Eichel, H. Kungl, and M. J. Hoffmann, Nanodomain structure of Pb [Zr 1− x Ti x] O 3 at its morphotropic phase boundary: Investigations from local to average structure, Physical Review B **75**, 184117 (2007).

[23] W.-F. Rao and Y. U. Wang, Microstructures of coherent phase decomposition near morphotropic phase boundary in lead zirconate titanate, Applied Physics Letters **91** (2007).

[24] K. Yan, S. Ren, M. Fang, and X. Ren, Crucial role of octahedral untilting R3m/P4mm morphotropic phase boundary in highly piezoelectric perovskite oxide, Acta Materialia **134**, 195 (2017).

[25] R. He, H. Wu, L. Zhang, X. Wang, F. Fu, S. Liu, and Z. Zhong, Structural phase transitions in SrTi O 3 from deep potential molecular dynamics, Physical Review B **105**, 064104 (2022).

[26] H. Wu, R. He, Y. Lu, and Z. Zhong, Large-scale atomistic simulation of quantum effects in SrTiO 3 from first principles, Physical Review B **106**, 224102 (2022).

[27] R. He, H. Xu, P. Yang, K. Chang, H. Wang, and Z. Zhong, Ferroelastic twin wall mediated ferro-flexoelectricity and bulk photovoltaic effect in SrTiO $_3$, arXiv preprint arXiv:2310.10130 (2023).

[28] R. He, B. Zhang, H. Wang, L. Li, T. Ping, G. Bauer, and Z. Zhong, Ultrafast switching dynamics of the ferroelectric order in stacking-engineered ferroelectrics, Acta Materialia, 119416 (2023).

[29] Y. Zhang, H. Wang, W. Chen, J. Zeng, L. Zhang, H. Wang, and E. Weinan, DP-GEN: A concurrent learning platform for the generation of reliable deep learning based potential energy models, Computer Physics Communications **253**, 107206 (2020).

[30] T. Wen, L. Zhang, H. Wang, E. Weinan, and D. J. Srolovitz, Deep potentials for materials science, Materials Futures **1**, 022601 (2022).

[31] H. Wang, L. Zhang, J. Han, and E. Weinan, DeePMD-kit: A deep learning package for many-body potential energy representation and molecular dynamics, Computer Physics Communications **228**, 178 (2018).

[32] D. Lu, W. Jiang, Y. Chen, L. Zhang, W. Jia, H. Wang, and M. Chen, DP compress: A model compression scheme for generating efficient deep potential models, Journal of chemical theory and computation **18**, 5559 (2022).

[33] J. Wu, Y. Zhang, L. Zhang, and S. Liu, Deep learning of accurate force field of ferroelectric HfO 2, Physical Review B **103**, 024108 (2021).

[34] R. He, H. Wu, Y. Lu, and Z. Zhong, Origin of negative thermal expansion and pressure-induced amorphization in zirconium tungstate from a machine-learning potential, Physical Review B **106**, 174101 (2022).

[35] J. Wu, J. Yang, Y.-J. Liu, D. Zhang, Y. Yang, Y. Zhang, L. Zhang, and S. Liu, Universal interatomic potential for perovskite oxides, Physical Review B **108**, L180104 (2023).

[36] L. Zhang, H. Wang, M. C. Muniz, A. Z. Panagiotopoulos, and R. Car, A deep potential model with long-range electrostatic interactions, The Journal of Chemical Physics **156** (2022).

[37] G. Kresse and J. Furthmüller, Efficient iterative schemes for ab initio total-energy calculations using a plane-wave basis set, Physical review B **54**, 11169 (1996).

[38] G. Kresse and J. Furthmüller, Efficiency of ab-initio total energy calculations for metals and semiconductors using a plane-wave basis set, Computational materials science **6**, 15 (1996).

[39] J. P. Perdew, K. Burke, and M. Ernzerhof, Generalized gradient approximation made simple, Physical review letters **77**, 3865 (1996).

[40] A. Togo, First-principles phonon calculations with phonopy and phono3py, Journal of the Physical Society of Japan **92**, 012001 (2023).

[41] A. Togo, L. Chaput, T. Tadano, and I. Tanaka, Implementation strategies in phonopy and phono3py, Journal of Physics: Condensed Matter (2023).

[42] S. Plimpton, Fast parallel algorithms for short-range molecular dynamics, Journal of computational physics **117**, 1 (1995).

[43] A. P. Thompson, H. M. Aktulga, R. Berger, D. S. Bolintineanu, W. M. Brown, P.

S. Crozier, P. J. in't Veld, A. Kohlmeyer, S. G. Moore, and T. D. Nguyen, LAMMPS- a flexible simulation tool for particle-based materials modeling at the atomic, meso, and continuum scales, Computer Physics Communications **271**, 108171 (2022).

[44] P. Ghosez and J. Junquera, First-principles modeling of ferroelectric oxide nanostructures, arXiv preprint cond-mat/0605299 (2006).

[45] O. Gindele, A. Kimmel, M. G. Cain, and D. Duffy, Shell Model force field for Lead Zirconate Titanate Pb (Zr1–x Ti x) O3, The Journal of Physical Chemistry C **119**, 17784 (2015).

[46] M. J. Haun, E. Furman, S. Jang, H. McKinstry, and L. Cross, Thermodynamic theory of PbTiO3, Journal of Applied Physics **62**, 3331 (1987).

[47] B. Noheda, D. Cox, G. Shirane, J. Gonzalo, L. Cross, and S. Park, A monoclinic ferroelectric phase in the Pb (Zr 1− x Ti x) O 3 solid solution, Applied physics letters **74**, 2059 (1999).

[48] B. Noheda, J. Gonzalo, L. Cross, R. Guo, S.-E. Park, D. Cox, and G. Shirane, Tetragonal-to-monoclinic phase transition in a ferroelectric perovskite: The structure of PbZr 0.52 Ti 0.48 O 3, Physical Review B **61**, 8687 (2000).

[49] R. Eitel and C. Randall, Octahedral tilt-suppression of ferroelectric domain wall dynamics and the associated piezoelectric activity in Pb (Zr, Ti) O 3, Physical Review B **75**, 094106 (2007).

[50] C. Michel, J.-M. Moreau, G. D. Achenbach, R. Gerson, and W. J. James, Atomic structures of two rhombohedral ferroelectric phases in the Pb (Zr, Ti) O3 solid solution series, Solid State Communications **7**, 865 (1969).

[51] D. Damjanovic, in *Piezoelectric and acoustic materials for transducer applications* (Springer, 2008), pp. 59.